\documentclass[useAMS,usenatbib,usegraphicx]{mn2e}

\newcommand{\I}{\mbox{\thinspace{\sc i}}}
\newcommand{\kms}{\mbox{\thinspace km\thinspace s$^{-1}$}}
\newcommand{\chisq}{\mbox{$\chi^2$}}
\newcommand{\ha}{\mbox{H$\,\alpha$}}
\newcommand{\hb}{\mbox{H$\,\beta$}}
\newcommand{\hg}{\mbox{H$\,\gamma$}}
\newcommand{\porb}{\mbox{$P_{\rm orb}$}}
\newcommand{\Il}{\mbox{\thinspace{\sc i}}$\,\lambda\,$}
\newcommand{\IIl}{\mbox{\thinspace{\sc ii}}$\,\lambda\,$}
\newcommand{\III}{\mbox{\thinspace{\sc iii}}}
\newcommand{\II}{\mbox{\thinspace{\sc ii}}}
\newcommand{\IV}{\mbox{\thinspace{\sc iv}}}
\newcommand{\msun}{\mbox{M$_{\odot}$}}

\title[Spiral waves and the secondary in V3885 Sgr]{Spiral waves and the secondary in the novalike variable, V3885 Sgr}
\author[L. E. Hartley et al.]{Louise E. Hartley$^{1}$, James R. Murray$^{1}\thanks{E-mail:
jmurray@swin.edu.au}$, Janet E. Drew$^{2}$, Knox S. Long$^{3}$\\
$^{1}$Mail number 31, Swinburne University of
Technology, PO Box 218, Hawthorn, Victoria 3122, Australia\\
$^{2}$Astrophysics Group, Imperial College London,
Blackett Laboratory, Prince Consort Road, London, SW7 2AZ, UK\\
$^{3}$Space Telescope Science Institute, 3700 San Martin Drive,
Baltimore, MD 21218, USA}

\begin{document}

\maketitle

\label{firstpage}

\begin{abstract}
We present seven nights' blue (4300--5000\AA) spectroscopy of the
novalike variable star, V3885~Sgr. The line spectrum shows a typical
combination of broad absorption and emission in \hg\ and \hb\ and
He\I, which is associated with the accretion disk. We also observe
anti-phased narrow emission, which we attribute to irradiation of
the secondary star. The He\IIl4686 and N\III--C\III--C\IV\ emission
lines are devoid of structure and are most likely formed in an
outflow. We measure radial velocity shifts in the absorption and
emission lines, from which we fit an orbital period of
$4.97126\pm0.00036$\,h. From the velocity semi-amplitudes of the
disk and companion star, we are able to constrain the binary mass
ratio to $q>0.7$.

The phase-folded spectra provide dense coverage of the entire
orbital cycle. Doppler tomograms of the hydrogen and He\I\ lines
reveal spiral structure in the accretion disk and the irradiated
donor star. We believe this is the first unambiguous detection of
spiral waves in a novalike variable.
\end{abstract}


%%%%%%%%%%%%%%%%%%%%%%%%%%%%%%%%%%%%%%
%%%%%%%%%%%%%%%%%%%%%%%%%%%%%%%%%%%%%%
\section{Introduction}
V3885 Sgr is a bright ($m_v=10.4$) Cataclysmic Variable, an
interacting binary system consisting of an white dwarf that accretes
matter from a low-mass companion. V3885 Sgr belongs to the novalike
subclass. The white dwarf accretes via a disk, yet the disk does not
undergo outbursts in luminosity, as in dwarf novae. Rather, it
persists in an outburst state with only small fluctuations in
luminosity of a few tenths of a magnitude.

In understanding the processes of disk accretion, the novalike systems
are valuable tools, because the accretion disk dominates the system
luminosity and is thermally stable.  Yet, it is precisely because the
disk continually outshines the component stars that it is exceedingly
difficult to measure the parameters of these binary systems with any
accuracy.

There have been several attempts to determine the orbital period of
V3885 Sgr. First \citet*{77cowley} derived an orbital period
\mbox{$P = 5.04\,$h} from radial velocity shifts in Balmer lines,
seen mainly in absorption.  Their observations seem to have been
gathered during four spells of around 2.5\,h each, spanning three
days.  Then \citet{85haug} used essentially the same spectroscopic
technique and emerged with a very different period of 6.216\,h.
There were other peaks in the power spectrum they derived that were
scarcely less pronounced, but none matched the period found by the
earlier study. Haug \& Drechsel gathered data on three occasions,
spanning 2 months, adding to a total of $\sim$12\,h observation.
Finally \citet{89metz} gave a period of 5.191\,h, without ever
presenting the experimental method clearly, yet this is the most
commonly quoted period.

V3885 Sgr has been observed repeatedly in the ultraviolet by the
{\it International Ultraviolet Explorer} ({\it IUE}\/)
\citep{92woods} and, more recently, by the {\it Hubble Space
Telescope} \citep{02hartley}. The ultraviolet spectrum is dominated
by strong wind-formed line profiles, but \citeauthor{02hartley}
found a surprising absence of short-timescale variability.
Ultimately our interpretation of the ultraviolet spectrum was
hampered by an inability to separate stochastic wind-formed features
from phase-dependent variability. Therefore, we set out to firmly
establish the orbital period of V3885~Sgr, by measuring radial
velocity shifts in the blue ($\sim4200-5100$\AA) line spectrum.

The observations are detailed in Section~\ref{s:obs}. In
Section~\ref{s:period} we present our measurement of the orbital
period. Then in Section~\ref{s:profiles} we examine the line
variability in more detail, using Doppler tomography to reveal the
secondary star and structure in the accretion disk.
%%%%%%%%%%%%%%%%%%%%%%%%%%%%%%%%%%%%%%%
%%%%%%%%%%%%%%%%%%%%%%%%%%%%%%%%%%%%%%%
\section{The observations}
\label{s:obs} We carried out seven nights of  observations, during
2003 September 2--9, at the South African Astronomical Observatory's
(SAAO) 1.9-m telescope. The telescope was equipped with the grating
spectrograph and SITe CCD. We used the number 4 grating, which gives
a first-order central wavelength of 4600\AA\ and range of 800\AA.
V3885~Sgr's optical brightness enabled us to take advantage of the
narrow, 200-\micron\ slit, which provides a spectral resolution of
$\sim1$\,\AA. With an average exposure time of 300\,s, this setup
yields a S/N$\sim25$ per exposure.

To ensure good wavelength calibration, CuAr arc exposures were taken
about every 15\,min. Also, comparison observations of the radial
velocity standard, HD187474, were made each night. In total, 300
useful spectra of V3885 Sgr were obtained over the seven nights and
the good weather conditions allowed us to record some data every
night.

\begin{figure}
\includegraphics[scale=0.33,angle=270]{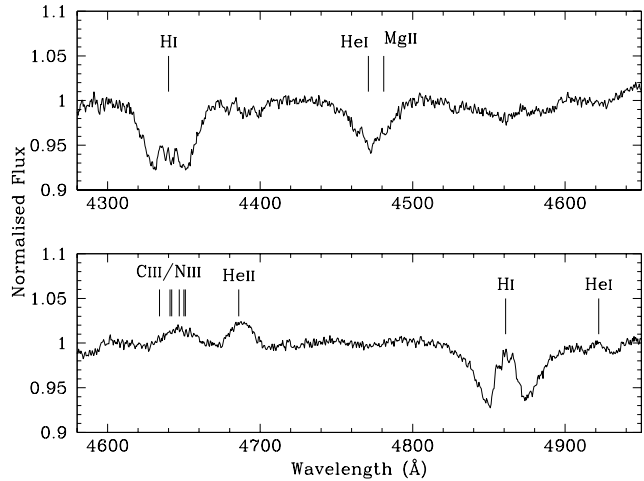}
\caption{The time-averaged spectrum of V3885 Sgr, as observed at the
SAAO in September 2003.\label{f:spec}}
\end{figure}

Fig.~\ref{f:spec} shows a normalised spectrum created from the sum of
all the exposures. The spectrum shows the same line features as seen
in \citet{85haug}, albeit at significantly higher S/N. The line
spectrum is typical of a mid-inclination novalike variable: the line
profiles are weak, consisting of broad Balmer and He\I\ absorption,
with superposed emission and weak emission in C\III/N\III\ and He\II\
(see Section~\ref{s:profiles}).

%%%%%%%%%%%%%%%%%%%%%%%%%%%%%%%%%%%%%%%%
%%%%%%%%%%%%%%%%%%%%%%%%%%%%%%%%%%%%%%%%
\section{The orbital period}
\label{s:period} Measuring radial velocity variations in the
absorption and emission lines made more difficult by the presence of
several components that make up each line. The hydrogen lines
consist of broad absorption, broad emission and narrow emission (see
Section~\ref{s:profiles}. As these components may be formed in
distinct parts of the system, each had to be isolated and its radial
velocity measured independently of the other components.

A least-squares method was used to fit Gaussians to each line
component. For each line component the FWHM was held constant,
depending on the line and the width of that feature. For the broad
absorption the central portion of the line containing the emission
components was masked out by eye and a Gaussian fitted to the line
wings. For the broad emission the narrow emission (where present)
and central double peak were masked out and two-component Gaussian
was fit to the emission and the underlying broad absorption (to
remove the latter). This fit was then subtracted from the line
profile in order to measure the narrow emission component.

\begin{figure}
\includegraphics[angle=0,scale=0.36]{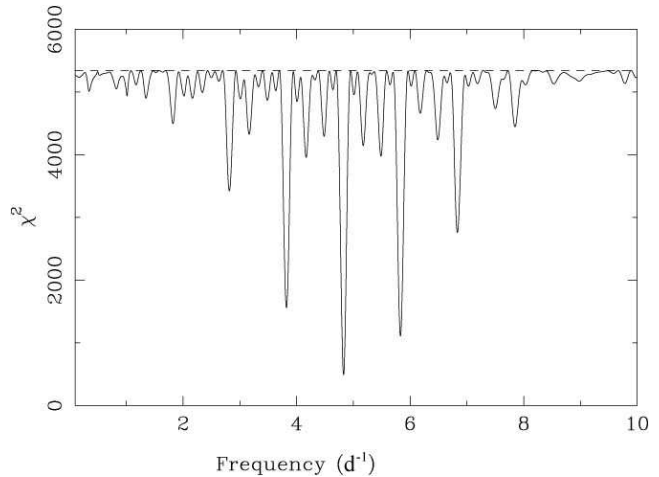}
\caption{A sample \chisq--frequency plot, for the H$\beta$ line.\label{f:pgram}}
\end{figure}

The best-fit radial velocity curve was found by folding the data on
a range of trial frequencies and finding the lowest \chisq\ sine
curve at each frequency (software supplied by T. Marsh, private
communication). A plot of \chisq\ against frequency (see
Fig.~\ref{f:pgram}) shows the most likely frequency. As expected for
nightly sampling, there are period aliases at $\pm1\,$d from the
strongest alias. However, the second strongest alias has a \chisq\
of more than double that of the strongest alias and so we can be
confident that the latter is the correct spectroscopic period.

\begin{figure*}
\includegraphics[scale=0.2,angle=270]{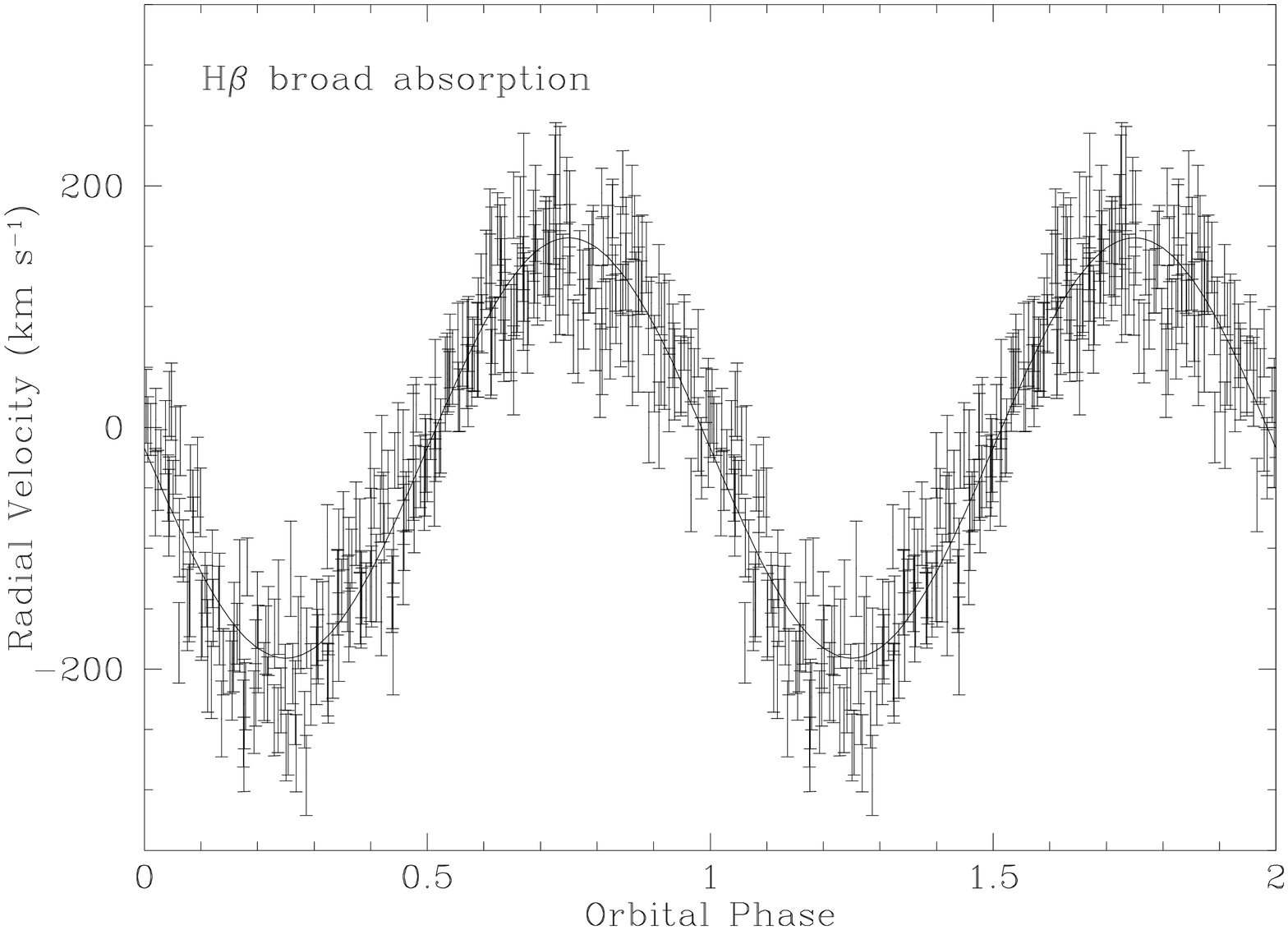}
\includegraphics[scale=0.2,angle=270]{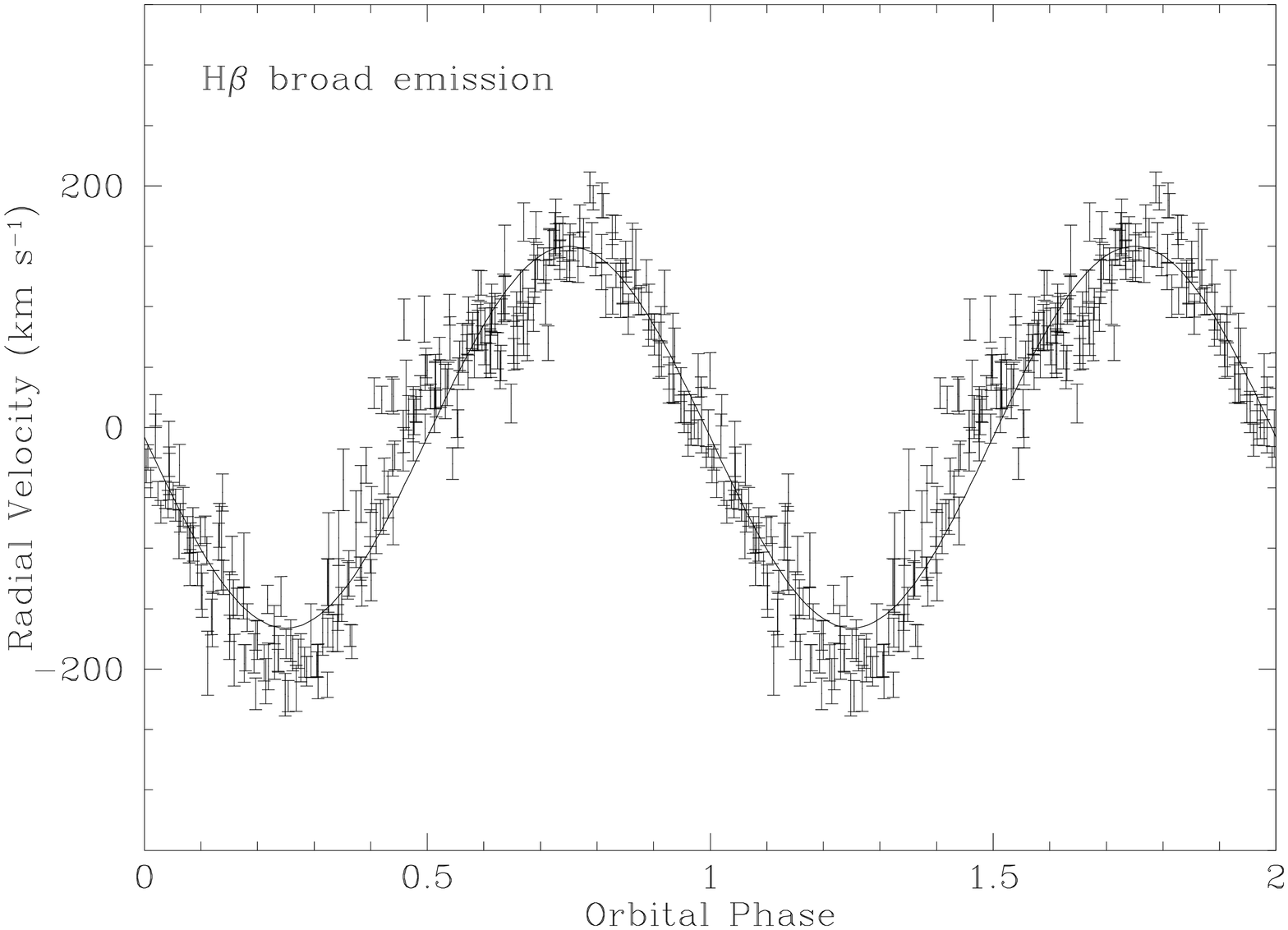}
\includegraphics[scale=0.2,angle=270]{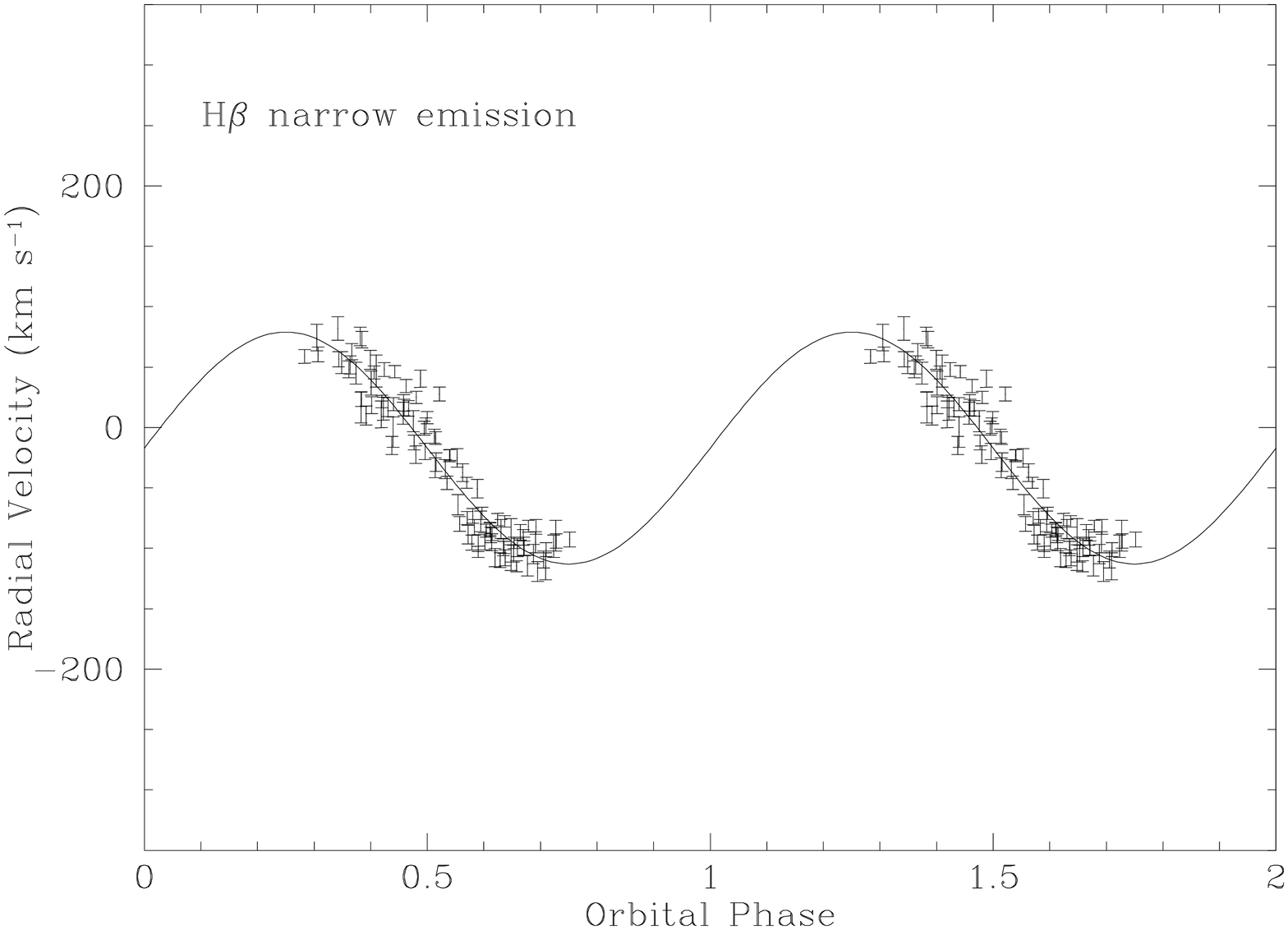}
\includegraphics[scale=0.2,angle=270]{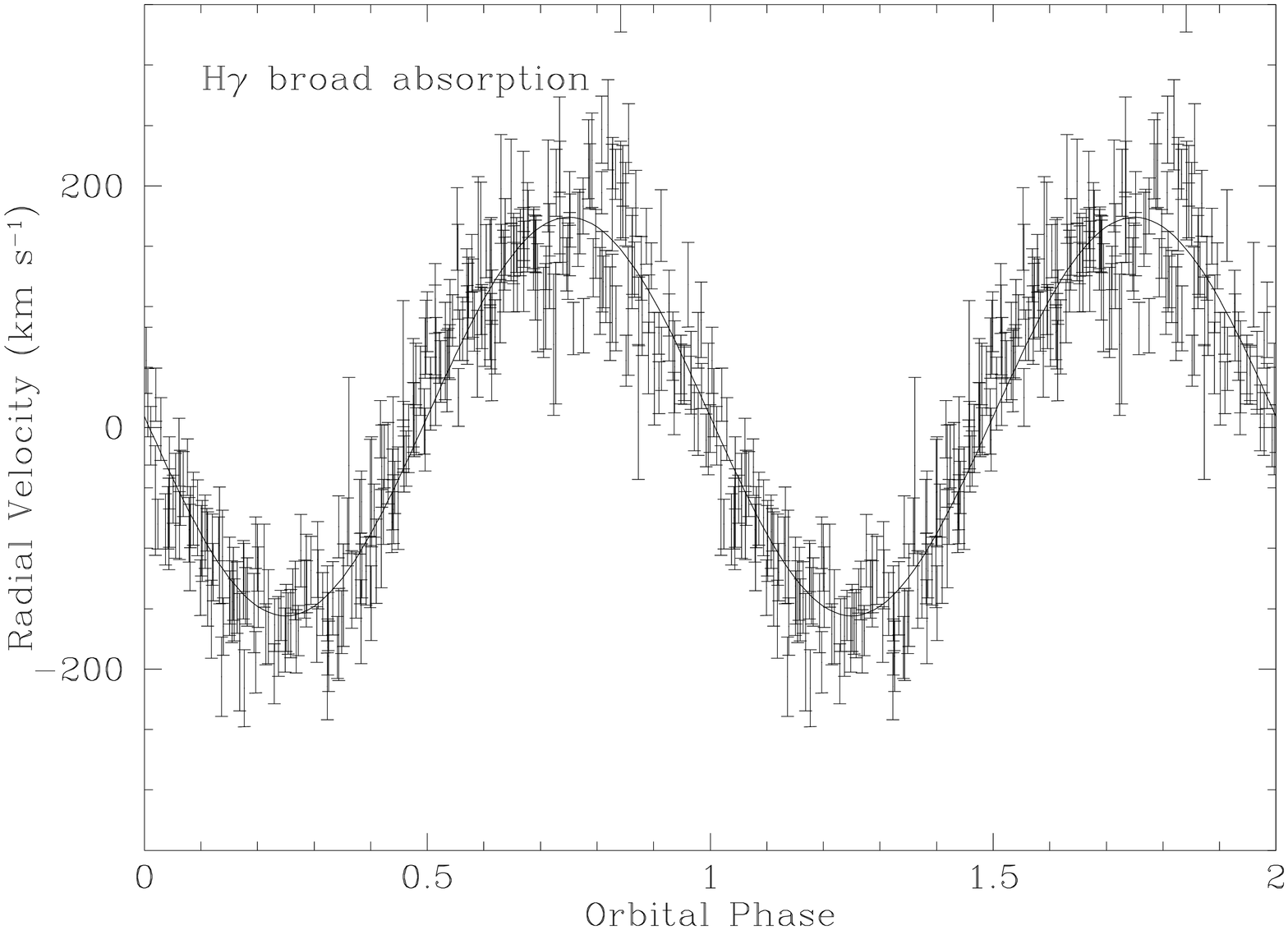}
\includegraphics[scale=0.2,angle=270]{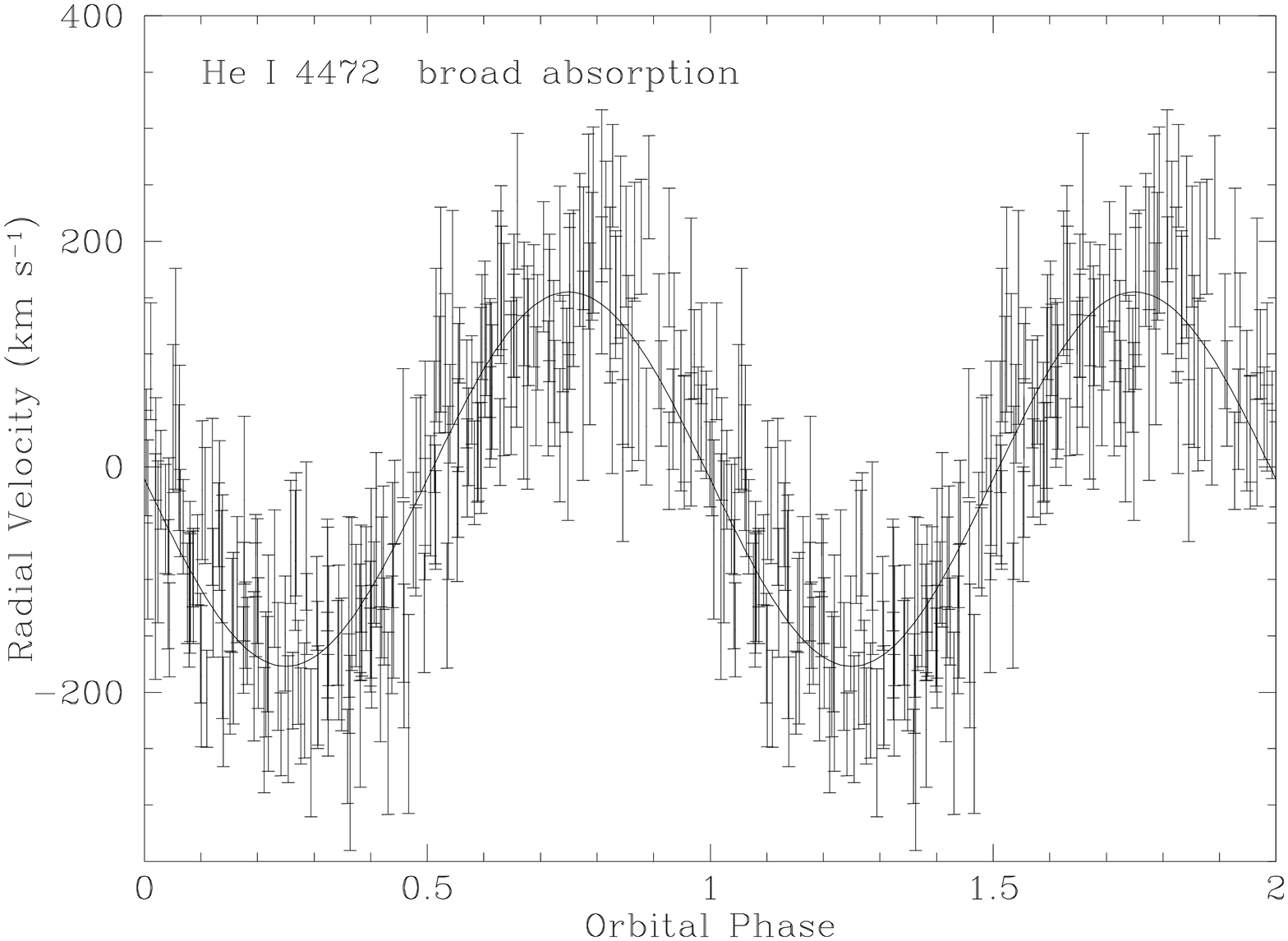}
\includegraphics[scale=0.2,angle=270]{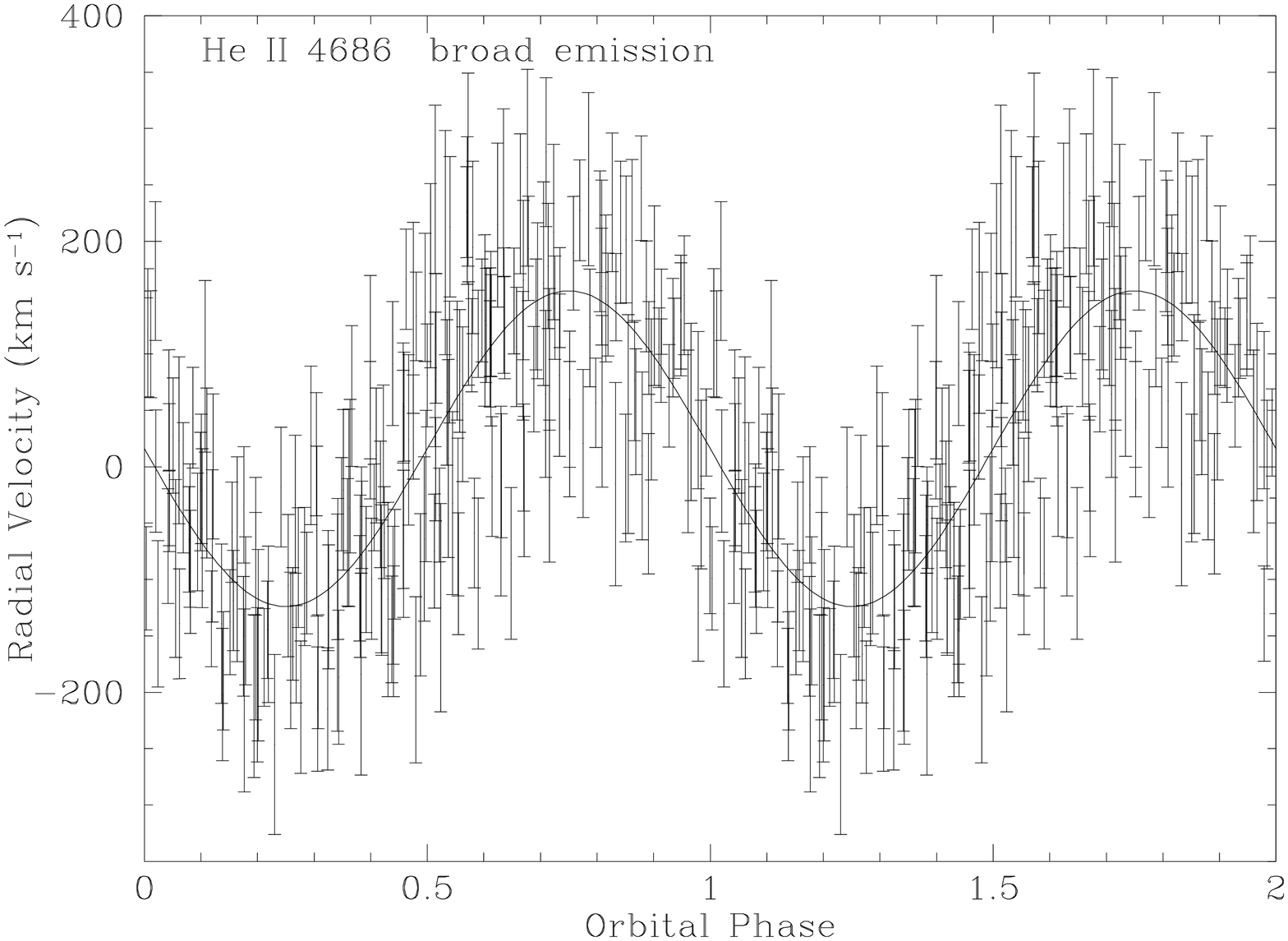}
\caption{Radial velocity curves of shifts in the absorption and
emission lines, folded on the orbital period. Top row, \hb: broad
absorption, broad emission and narrow emission. Bottom row: \hg\ broad
absorption, He\Il4472 broad absorption and He\III4686 broad
emission.\label{f:shifts}}
\end{figure*}

The Gaussian fit technique worked well for the \hb\ line. However,
we found that the \hg\ broad absorption is not well fit by a
Gaussian; it has steeper absorption wings, which descend abruptly
from the continuum. To measure shifts in this line, we instead used
a cross-correlation technique. First a template spectrum was
created. The spectra were assigned radial velocities according to
that measured for \hb\ and then the corresponding radial velocities
were subtracted from the wavelength scale of each spectrum. These
spectra were summed to create a radial-velocity corrected template.
After masking out all but the wings of the \hg\ line, each spectrum
was cross-correlated with this template. This method was also used
for the He\Il4472 absorption and He\IIl4686 emission lines. Results
of the radial-velocity measurements are listed in
Table~\ref{t:periods}. Fig.~\ref{f:shifts} shows the radial velocity
points, folded at their least-\chisq\ period for each of the
H$\beta$ line features.

\begin{table*}
\begin{tabular}{@{}lcccccc}
\hline Line    &$K$&$\gamma$   &$P_{\rm orb}$ &$T_0$ &N. Points
&$\chi^2_
\nu$\\
    &(\kms) &(\kms) &(h)&(MHJD)&\\
\hline
%H$\beta$ broad absorption   &174(5) &-17(3) &0.20715(08)    &4.9716(19) &52887.86076(8)&295&1.67\\
%H$\beta$ broad emission     &158(4) &-8(3) &0.20720(04)
%&4.9728(10) &52887.85840(13)    &280    &5.57\\
H$\beta$ broad absorption and emission  &163(2) &-10(2)
%&0.20718(04)
&4.9724(10) &52887.85890(04)&575&3.68\\
H$\beta$ narrow emission    &97(2)     &-19(5)
%&0.20717(05)
&4.9721(12) &52887.9659(20) &105&8.05\\
H$\gamma$ broad absorption  &165(6)    &$^a$
%&0.20707(08)
&4.9697(19) &52887.85911(08)&285&1.70\\
He\Il4472 broad absorption  &166(10)   &$^a$
%&0.20701(16)
&4.9682(38) &52887.8654(19) &225&1.15\\
He\IIl4686 broad emission   &135(10)&$^a$
%&0.20721(23)
&4.9730(55) &52887.8595(24)&224&2.03\\
\hline
\end{tabular}
\\$^a$Measured from cross correlation of
template spectrum, so no direct measurement of $\gamma$.
\caption{Parameters of radial-velocity fits to spectral lines in the
blue spectrum of V3885~Sgr. Errors in parentheses are 2$\sigma$
limits.\label{t:periods}}
\end{table*}

The periods derived from the measured features agree to within the
errors and a weighted mean gives an orbital period of
$4.97126\pm0.00036$\,h. To within the errors the \hb\
 narrow emission moves in exact anti-phase to the other lines. Setting aside the
\hb\ narrow emission, we obtain an average zero-phase of
$52887.8602\pm0.0052\,$MHJD, which refers to the epoch of
red-to-blue crossing of the line features. If we assume that these
features are formed in or above the accretion disk, this is inferior
conjunction of the secondary star.

We find good agreement between the $K$-velocities obtained from the
\hb, \hg\ and He\Il4472 lines, with an average $K=166\pm6$\kms.
He\IIl4686 is fit by a lower $K-$-velocity of $135\pm10$\kms. A close
inspection of a trailed spectrum reveals a very slight weakening of
the short-wavelength edge of the He\II\ line at around phase 0.25.
This may account for the lower measured $K$-velocity. To improve the
cross-correlation in this range of the spectrum, we attempted to
measure He\IIl4686 together with the 4630--4660\AA\ blend, however
the latter was too weak in the individual spectra to give a reliable
measurement.
%%%%%%%%%%%%%%%%%%%%%%%%%%%%%%%%%%%%%%%%%%%%%
%%%%%%%%%%%%%%%%%%%%%%%%%%%%%%%%%%%%%%%%%%%%%
\section{line profiles}
\label{s:profiles}

We find the optical spectrum of V3885~Sgr unchanged since the 1982
observations of \citet{85haug}. The Balmer lines (\hb\ and \hg) show
broad absorption, of a full width at the continuum $\sim5000$\kms,
and complex emission cores (see Section~\ref{ss:diskem}). He\Il4472
is also seen as broad absorption, with some structured emission near
line centre. Other He\I\ lines at 4387, 4713 and 4922\AA\ can be
seen as weak absorption/emission lines.

The broad H\I\ and He\I\ absorption line profiles remain very constant
in shape over the orbital period and their width suggests they are
formed in the optically thick accretion disk.

Weak emission lines of FWZI$\simeq2000\,\kms$ are present in the
carbon/nitrogen blend at 4650\,\AA\ and He\IIl4686 line. This broad
emission appears devoid of the emission structure associated with
the broad absorption lines. The 4630--4660\AA\ blend is commonly
thought to be comprised of C\III\ and N\III\ emission. However, a
Gaussian fit that imposes the correct relative multiplet wavelengths
of the N\III, C\III\ and C\IV\ components indicates that the
line profile is better emulated with a non-negligible C\IV\ contribution.

The He\IIl4686 emission profile probably forms mainly in an outflow.
It is no wider in velocity terms than the UV wind-dominated
He\IIl1640 absorption \citep[see][]{02hartley} and the lower level
of the 4686\AA\ line is the upper level of He\IIl1640, so He\IIl4686
can be recombination dominated even when He\IIl1640 is in
absorption. Furthermore any outflow may degrade the radiation field
irradiating the outer disk, eliminating He$^+$ ionising photons. So
we need not see any He\IIl4686 structure associated with secondary
or disc warps. Also, the He\IIl4686 emission shifts velocity in
phase with the disk-formed \hb\ line (Table~\ref{t:periods} and
Figure~\ref{f:shifts}), but by a slightly smaller amount
($K$=135\kms, compared to 166\kms). We suspect either blueshifted
absorption and/or opacity effects cause us to detect only flux from
the nearside of the white dwarf, thus distorting the velocity
profile. The C\III--N\III--C\IV\ line is too weak to measure its
velocity shifts. However, its similarity to He\IIl4686 indicate that
this feature is also most likely to be formed in the same region.

%%%%%%%%%%%%%%%%%%%%%%%%%%%%%%%%%%%%%%%%%%%%%
\subsection{Structure in the accretion disk}
\label{ss:diskem} 
\begin{figure}
\includegraphics[scale=0.55]{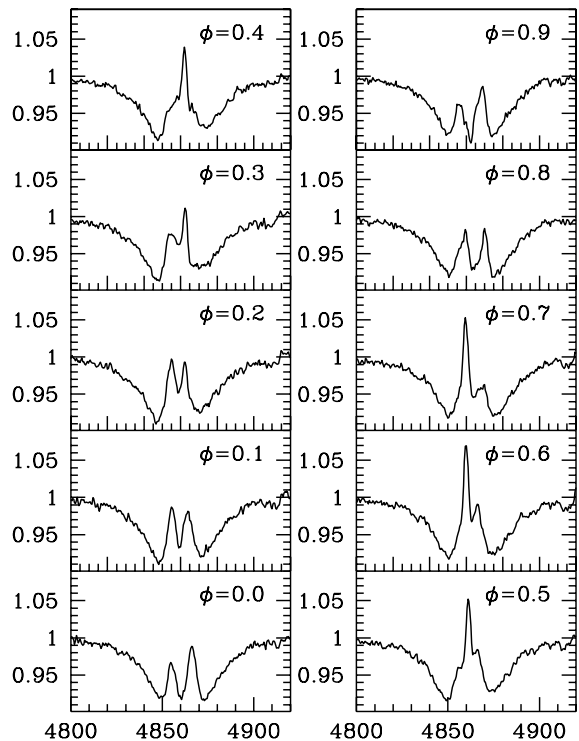}
\caption{The variation of the \hb\ profile over an orbital cycle.
The spectra are phase-binned to 0.1$P_{\rm orb}$. The flux scale is
relative intensity.\label{f:hbeta}}
\end{figure}

Fig.~\ref{f:hbeta} shows the variation of the \hb\
line profile over an orbital cycle. The \hg\ and He\Il4472 lines
undergo similar changes in shape, although in those lines, the ratio
of emission to absorption is smaller. The complex time-variable
emission, which is superposed onto the broad absorption can be
deconvolved into two components: broad emission of
FWHM$\simeq$700\kms\ and narrow emission of FWHM$\simeq$150--200\kms
(see~Section~\ref{ss:irad}). At those phases when the narrow
emission is not seen ($\phi\simeq$0.8--0.2) the broad emission can
in fact be seen to be double-peaked and composed of two emission
profiles of FWHM$\simeq$400\kms, with peaks separated by about
600\kms.

The dense phase-coverage of our spectra lends them well to
Doppler-mapping techniques \citep{88marsh}. Trailed spectra were
made by normalising each spectrum by the time-average and then
stacking the resultant spectra. These trailed spectra were the input
for the Doppler mapping. We succeeded in making Doppler maps from
trailed spectra of the \hg, \hb, He\Il4371 and He\Il4922 lines.
Attempts were made to create a Doppler map from He\IIl4686, but the
map did not converge on any particular distribution of He\II,
probably because of the low S/N ratio of this line.

\begin{figure*}
\includegraphics[scale=0.68]{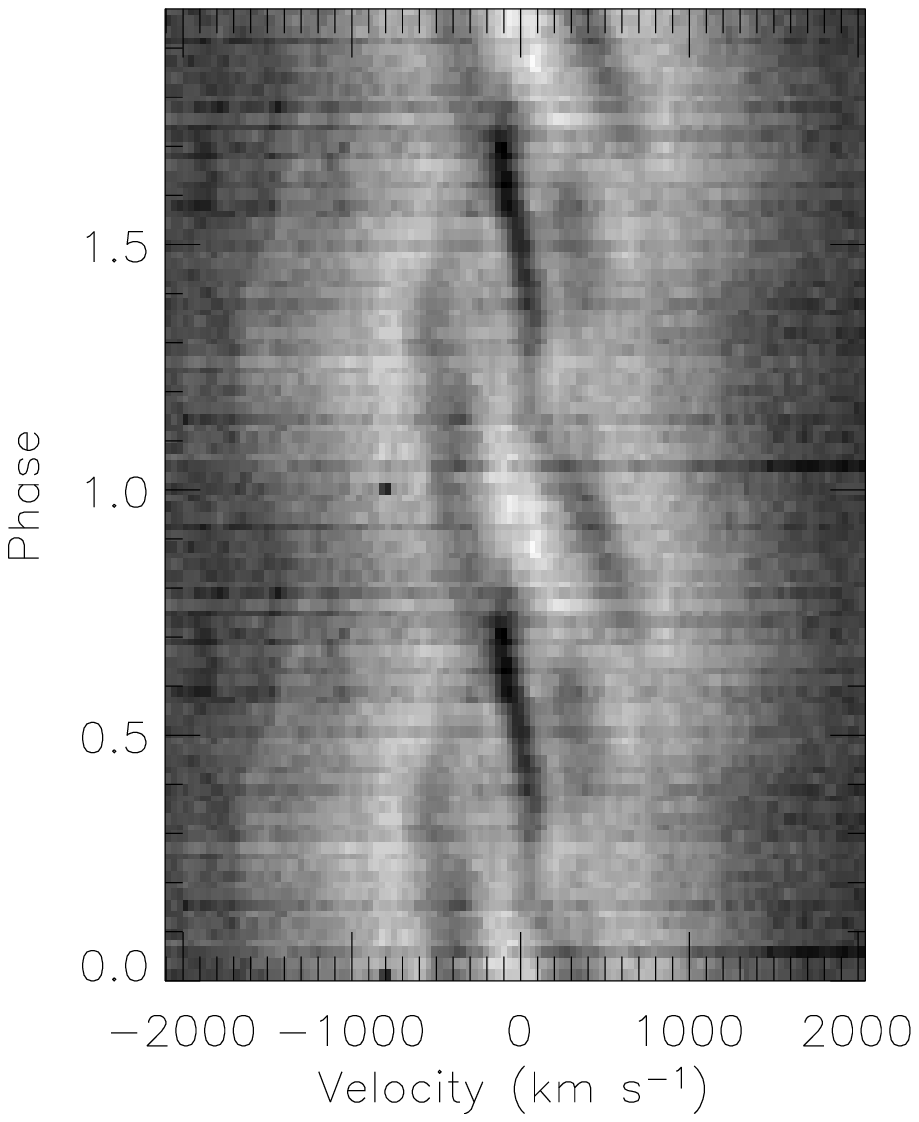}
\includegraphics[scale=0.52]{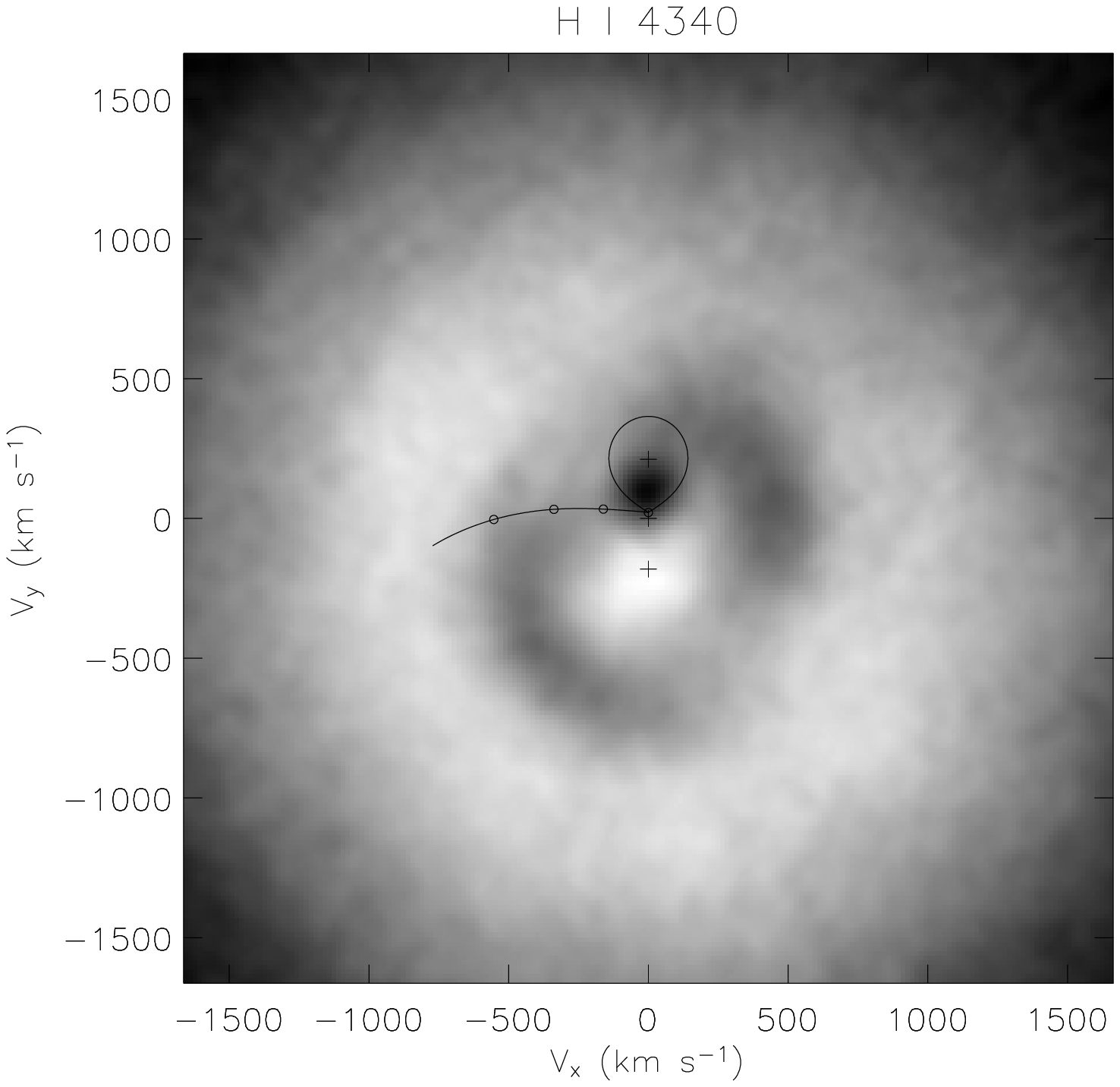}
\caption{Trailed spectrum and Doppler tomogram of the \hg\
line. Marked on the map are the locations of the primary, secondary
and system centres of mass; the Roche lobe of the secondary; and the
accretion stream from the secondary. These were calculated for input
parameters of $M_1$=0.7\msun, $M_2$=0.6\msun, and $i=65\degr$. The
flux scale is from white to black (i.e. strongest absorption in white,
emission in black).\label{f:hgmap}}
\end{figure*}

\begin{figure*}
\includegraphics[scale=0.68]{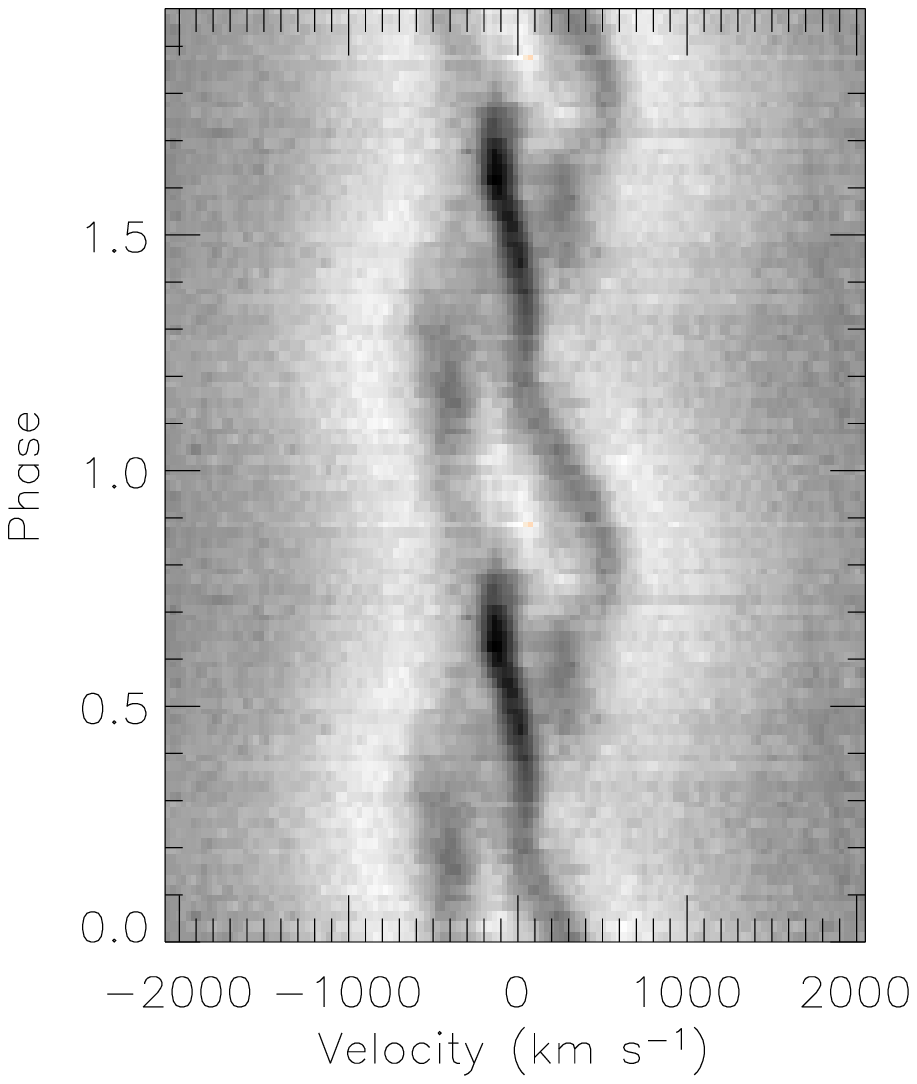}
\includegraphics[scale=0.52]{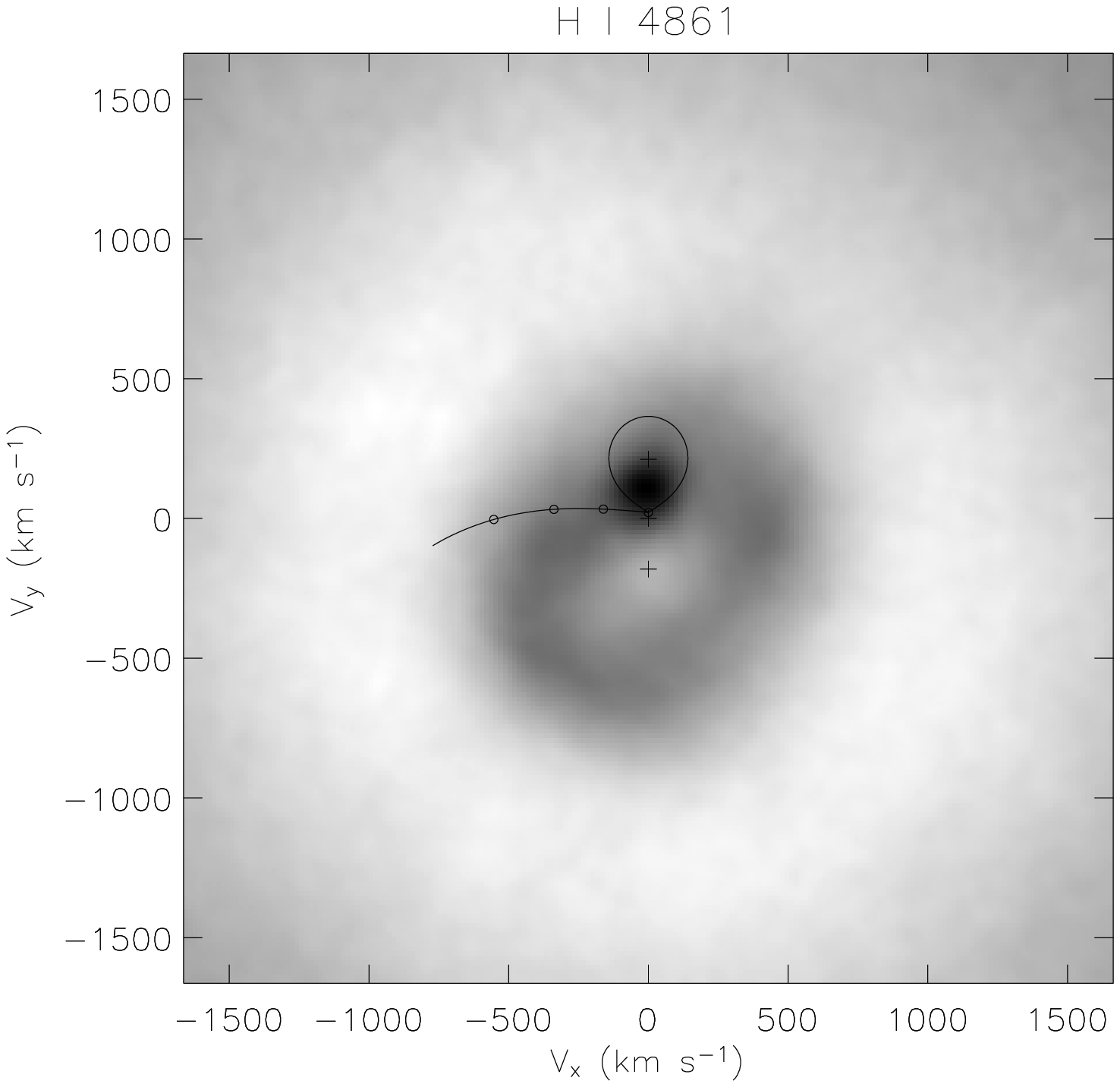}
\caption{Trailed spectrum and Doppler tomogram of the \hb\
line. Markings are as for Fig.~\ref{f:hgmap}. \label{f:hbmap}}
\end{figure*}

\begin{figure*}
\includegraphics[scale=0.68]{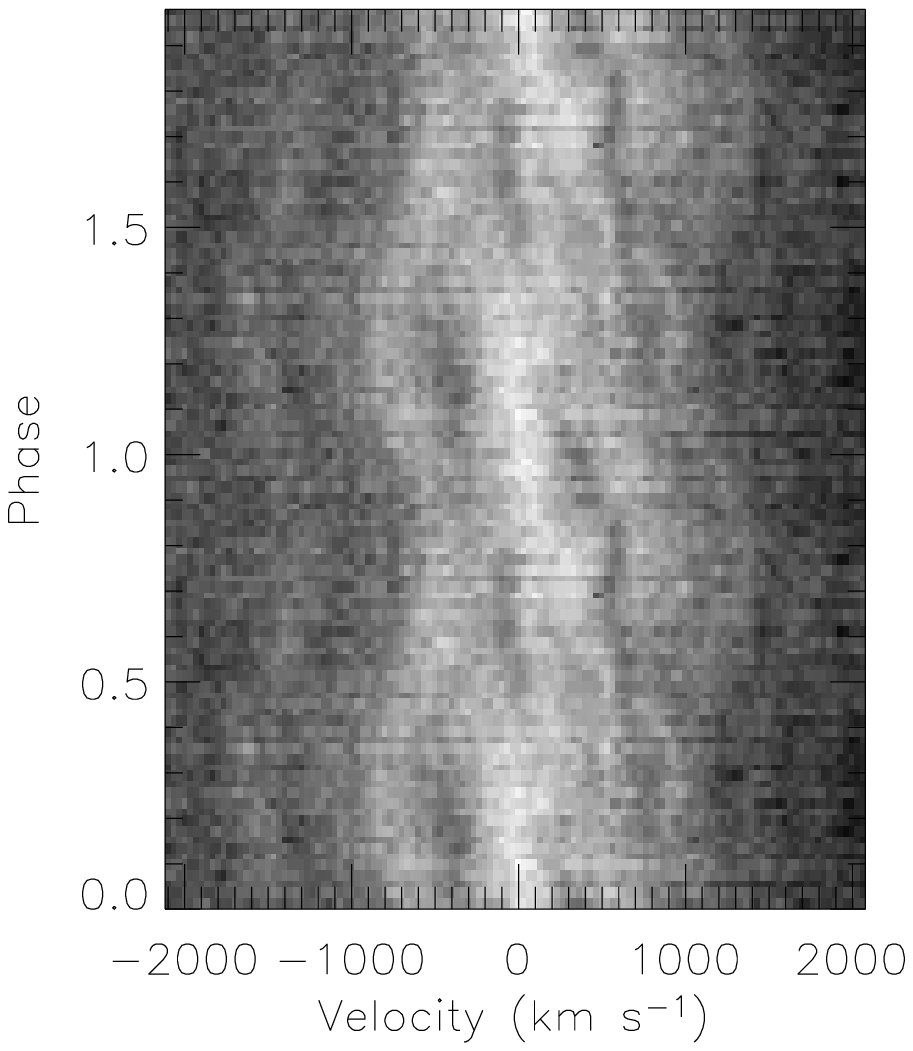}
\includegraphics[scale=0.52]{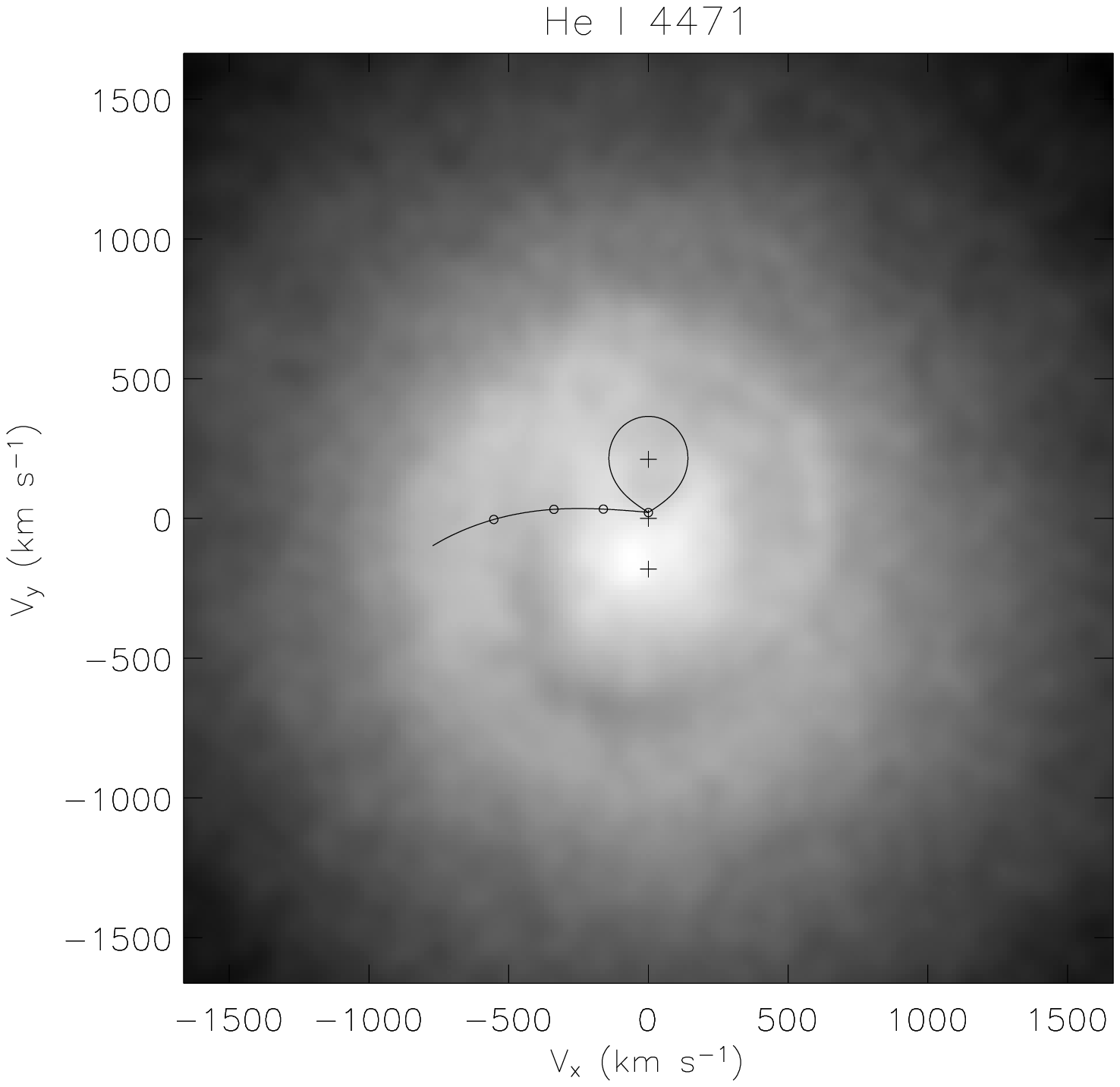}
\caption{Trailed spectrum and Doppler tomogram of the He\Il4472
line. Markings are as for Fig.~\ref{f:hgmap}. \label{f:heimap}}
\end{figure*}

The maps were created using an entropy maximisation technique (T.
Marsh, private communication). These are shown in
Figs.~\ref{f:hgmap} to \ref{f:heimap}. The maps require the systemic
($\gamma$) velocity as an input parameter. We used the \hb\
$\gamma$\ velocity of -19\kms. The Roche lobe, centres of mass and
accretion stream, which are overplotted on the maps are for
arbitrary input parameters of $M_1$=0.7\msun, $M_2$=0.6\msun, and
$i=65\degr$. These were chosed as they are consistent with our
estimated system parameters (see Section~\ref{s:param}) and appear
to fit reasonably well to the position of the two stars in the maps.

In the H\I\ maps (Figs.~\ref{f:hgmap} and \ref{f:hbmap}) the
accretion disk itself can be seen in absorption (particularly well
in the \hg\ and He\Il4472 lines maps). The line emission is clearly
separated into the two components: emission from the irradiated side
of the secondary and from the disk. The disk emission is not
azimuthally symmetric, but appears as a bright elliptical ring. The
ring is similar to the signature generated by spiral structure in
the disk, as seen in the simulations of, e.g. \citet{99steeghs}.
Arguably the emission extends further along the spiral arms in \hg\
than in \hb, particularly the arm which curls around to the upper
right of the secondary emission in the maps. The spiral structure of
the emission explains the deviation from a pure Gaussian of the line
profile, which leads to distortion of the radial velocity curve in
Fig.~\ref{f:shifts}.

The He\I\ lines are much weaker than the H lines, yet we still
manage to extract some information about the distribution of He\I\
in the accretion disk by creating a Doppler map from He\Il4472. This
line is blended with a weak Mg\IIl4481 line, which can be seen in
the trailed spectrum as periodic emission at the redward edge of the
absorption profile. We were able to include this component as a
blend in the Doppler tomogram, but were forced to estimate its
strength relative to He\Il4472. However, we can see in
Fig.~\ref{f:heimap} that He\Il4472 follows the same spiral structure
as \hg\ and \hb. It is also just possible to discern a faint
emission centred on the secondary, but it is much weaker than in the
H\I\ maps. The emission is, however, clearly present in the trailed
spectrum. The relative strength of the emission, compared to
absorption, is significantly weaker than for the hydrogen lines.

\begin{figure}
\includegraphics[scale=0.68]{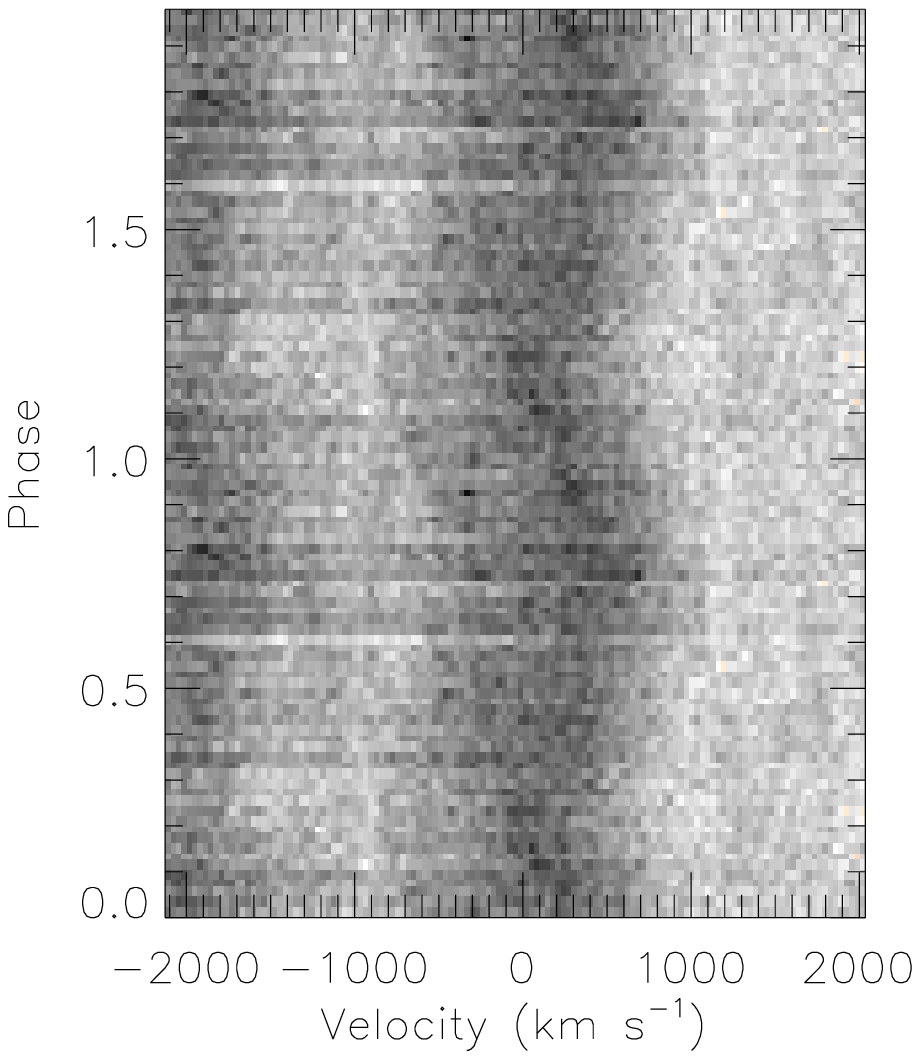}
\caption{Trailed spectrum of the He\Il4686 line.
\label{f:heiitrail}}
\end{figure}

\begin{figure}
\includegraphics[scale=0.68]{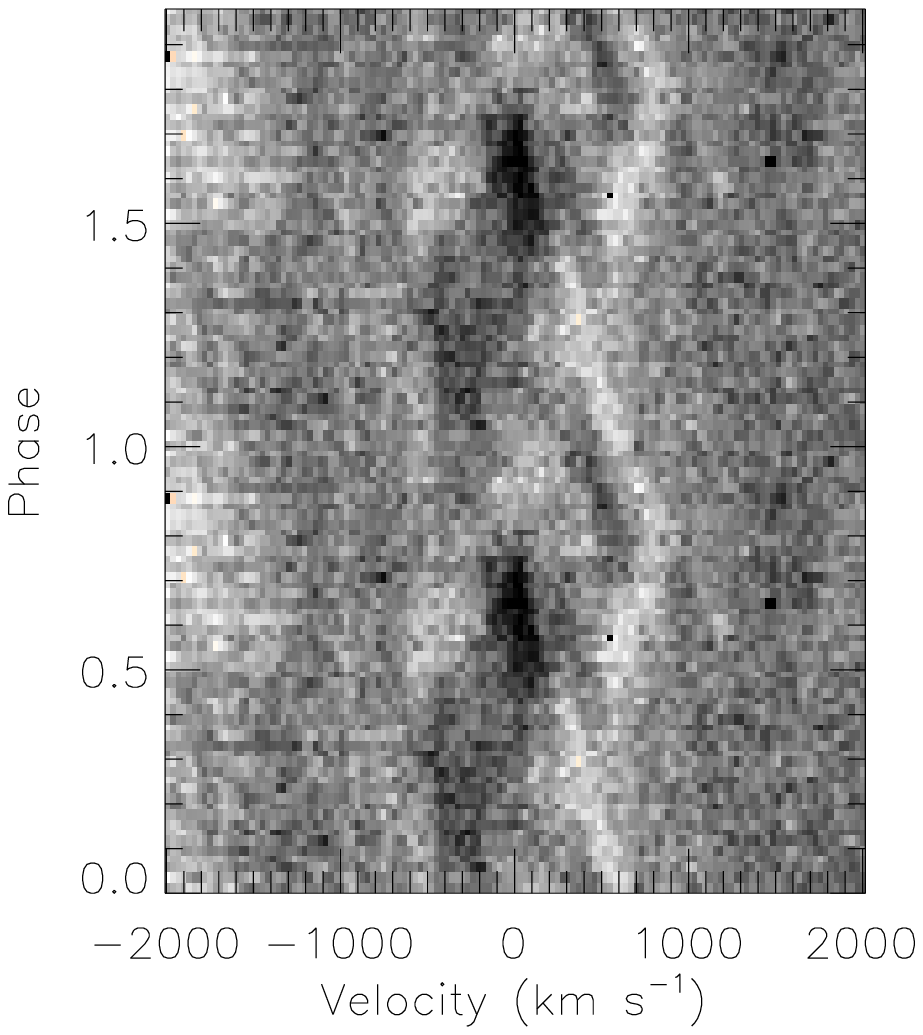}
\caption{Trailed spectrum of the He\Il4922 line. \label{f:hei4map}}
\end{figure}

The low S/N ratio of the He\Il4922 line precludes us from
constructing a reliable Doppler map, likewise for He\IIl4686.
However we have plotted their phase-folded trailed spectra in
Figs.~\ref{f:heiitrail} and \ref{f:hei4map}. From the trailed
spectrum of He\Il4922 we can discern the bright narrow emission that
we have attributed to the irradiated secondary. We also see the
S-wave structure, which is revealed as spiral arms in the H\I\ maps.
Less structure is visible in He\IIl4686, although we can just make
out a darker structure (corresponding to emission) moving bluewards
at phase 0.8 to 1.2, which in the \hb\ and \hg\ trailed spectra is
associated with the spiral seen in the upper-right quadrant of the
Doppler maps.

%%%%%%%%%%%%%%%%%%%%%%%%%%%%%%%%%%%%%%%%%%%%%
\subsection{Irradiation of the secondary star}
\label{ss:irad} 
\begin{figure}
\includegraphics[scale=0.43]{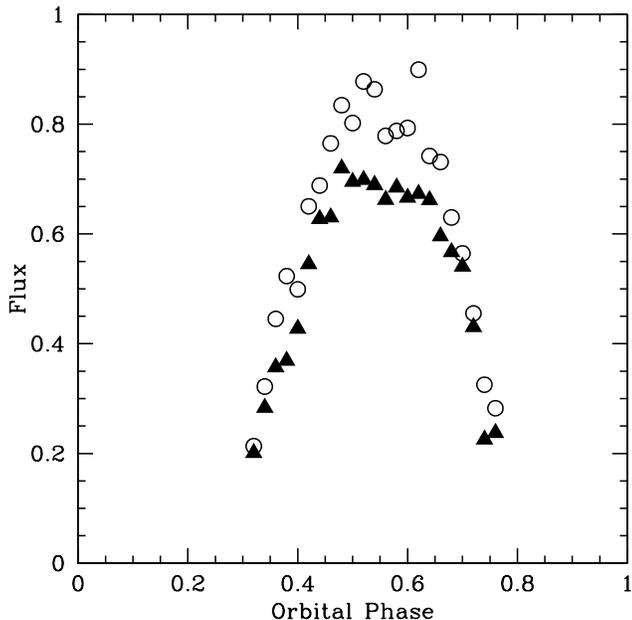}
\caption{Variation in the flux of the narrow emission feature over
an orbital cycle, showing H$\beta$ (open circles) and H$\gamma$
(filled triangles). The two line fluxes have a FWHM of
$\sim0.4$.\label{f:nemis}}
\end{figure}

To measure the strength of the narrow emission
requires that it first be isolated from the complex underlying line
structure. To do this we first radial-velocity corrected each
individual spectrum to remove the WD orbital motion. From the corrected
spectra a time-average was created, which was used to normalise each
radial-velocity corrected spectrum. Finally, these normalised spectra
were phase binned to a resolution of 0.02$P_{\rm orb}$. As the
underlying broad absorption and double-peaked emission remains
fairly constant over the orbital cycle, the only feature that varies
significantly in the normalised spectra is the narrow emission, thus
we were able to measure the variation in the narrow emission flux
(see Fig.~\ref{f:nemis}).

The narrow emission is present in the \hb\ and \hg\ lines from phase
0.32 to 0.76, with peak flux centred around phase 0.55. It has a
FWHM of around 150--200\kms. Narrow emission can also be
distinguished in He\Il4472 and $\lambda$ 4922, but appears later at
phase 0.50 to 0.76, with a peak at phase 0.55. However, the S/N of
this narrow component is very low and so hard to distinguish from
the other line components. There is no evidence of narrow emission
in He\IIl4686, nor in the C\III--C\IV--N\III\ complex.

The radial-velocity curve of the narrow \hb\ emission is antiphased
to that of the broad absorption. So the narrow emission most
probably emanates from an irradiated region on the the secondary
star facing the WD. This would also explain the disappearance of the
emission when irradiated side of the secondary rotates away from us
(around phase 0). The narrow emission peaks in strength at phase
0.55. We would expect irradiation of the secondary solely by the WD
and inner disk to give rise to emission appearing symmetrically
around phase 0.5, whereas the peak emission lags the orbital motion
by 0.05\porb\ (its appearance and disappearance also shows roughly
the same phase lag). This suggests either that the irradiating
source is offset from the line of centres of the binary system,
consistent with a hot-spot formed at the point of impact of the
accretion stream on the outer disk, or that the the irradiation due
to the WD is impeded by, for example, vertical disc structure.

%%%%%%%%%%%%%%%%%%%%%%%%%%%%%%%%%%%%%%%%%%%%%
%%%%%%%%%%%%%%%%%%%%%%%%%%%%%%%%%%%%%%%%%%%%%
\section{Discussion}
We have measured the spectroscopic orbital period of V3885 Sgr at
$4.97126\pm0.00036$\,h. The emission and absorption line profiles
and the Doppler maps created from them have
revealed:\begin{itemize}\item{the irradiated inner surface of the
companion star}\item{the accretion disc, showing spiral structure,
but no sign of any bright spot at the outer edge}.\end{itemize}

%%%%%%%%%%%%%%%%%%%%%%%%%%%%%%%%%%%%%%%%%%%%%
\subsection{Irradiation of the secondary}
The narrow line emission associated with the irradiated companion is
seen in the H lines, lasting from \mbox{$\phi=0.32$--0.76} and peaking
at $\phi=0.55$. In He\I\ the same narrow emission appears later at
$\phi\simeq0.5$. It is not seen in He\IIl4686, nor the nearby
N\III--C\III--C\IV\ blend. The anti-phasing of the narrow emission
component to the disk-formed absorption (Fig.~\ref{f:shifts}) places
it on the Doppler maps (Figs~\ref{f:hgmap} to
\ref{f:heimap}) at the inner edge of the companion star, suggesting
that this area is being heated by some irradiating source.

To better understand the emission mechanisms at work in V3885 Sgr,
we find it helpful to continue our comparison with the other bright
nova-like variable, IX Vel \citep[see][]{02hartley}. A detailed
spectral analysis of IX~Vel, covering the same spectral range as in
this paper (as well as \ha), is presented by \citet[hereafter
BT90]{90beuermann}. In IX~Vel the same narrow H$\beta$ emission is
seen to peak in strength at $\phi=0.55$. However, it is visible over
the whole orbital cycle (although BT90 admit that this could be due
to confusion between the narrow and broad components). In IX~Vel,
the variation of line flux with phase has a narrower FWHM of 0.35,
compared to 0.4 for V3885~Sgr (Fig.~\ref{f:nemis}).

BT90 model the variation of line flux produced by irradiating the
secondary from a bright spot at the outer edge of the disk and from
the WD (or boundary layer). They conclude that, to explain the
offset in peak flux from phase 0.5 to phase 0.55, some contribution
from the bright spot to the irradiating flux is probably required.
However, if we were to accept the bright spot as the irradiating
source for V3885 Sgr, we would expect it to be a visible component
in the Doppler maps \citep[e.g.][]{96har}.

In explaining the irradiation of the secondary during an outburst of
the dwarf nova IP Peg, \citet{90marsh} discount the bright spot as
the ionising source because the secondary is not seen to be
irradiated in quiescence, when the bright spot should presumably be
relatively brighter. They propose the boundary layer as the
irradiating source from which photons in the hydrogen, but not
helium Lyman continuum reach the secondary star (in common with
V3885 Sgr, IP Peg shows irradiation of the secondary in \hb\ and
\hg\ but not He\IIl4686).

We consider that a more plausible alternative to bright-spot
irradiation is an irradiating source located close to the WD, either
at the inner disk or boundary layer. In this scenario absorption by
vertical structure in the accretion disk distorts the radiation
field, causing the observed asymmetric irradiation pattern. Likely
sources for such structure are the spiral waves themselves, or the
overflow of the accretion stream across the surface of the disk. As
the resolution of our maps is at present, not sufficient to reveal
the distribution of either of these features in the inner disk
region, it is impossible to speculate any further.

%%%%%%%%%%%%%%%%%%%%%%%%%%%%%%%%%%%%%%%%%%%%%
\subsection{System parameters}
\label{s:param}

Estimates of the binary parameters of V3885~Sgr have been scarce
hitherto. However, it is possible to use our measured $K$-velocities
to constrain the component masses and system inclination.

\begin{figure}
\includegraphics[angle=0,scale=0.38]{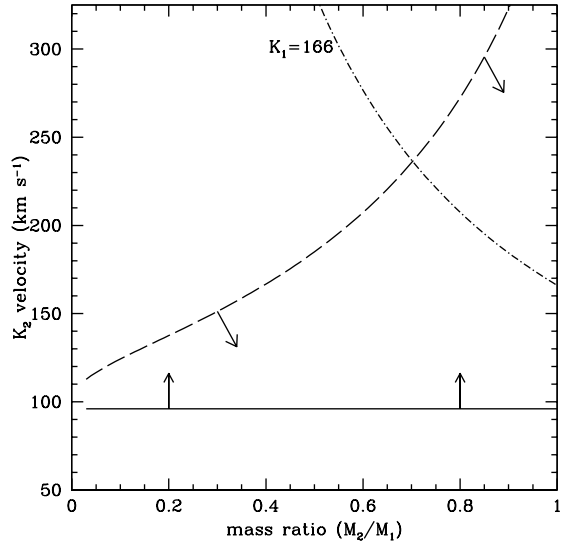}
\caption{Constraints to the value of $K_2$ as a function of mass
ratio \citep[as in Fig. 5 of][]{02steeghs}. The horizontal solid
line is the lower limit on $K_2$, given our measured $K'_2$. The
dashed line gives an upper limit on $K_2$, given that the velocity
of the $L_1$ point may not exceed $K'_2$. The dot-dashed lines
indicate K2 given the relationship $K_2=K_1/q$.\label{f:K2}}
\end{figure}

In Fig.~\ref{f:K2} we see the limits placed on the mass ratio, $q$,
by our measured values of $K_1$ (the WD $K$-velocity) and $K'_2$
(the $K$-velocity of the irradiated inner surface of the secondary).
These allow us to constrain $K_2$ (the secondary $K$-velocity) in
several ways: 1) it must be larger than $K'_2$ (solid line); 2)
$K'_2$ cannot be less than the $K$-velocity at the inner Lagrangian
point (dashed line); 3) $K_2=K_1/q$ for $K_1$=166\kms (dot-dashed line).

From Fig.~\ref{f:K2} we see that a mass ratio $q\ga0.7$ is required
to reconcile the two $K$-velocity measurements. We get an
independent estimate of the WD mass with the relation
$M_2\simeq0.11P_{\rm orb}$(h) \citep{88king}. For V3885~Sgr, $P_{\rm
orb}\simeq5$\,h, which translates to a secondary mass of
$\simeq0.55$\,\msun. For $q\ga0.7$ this requires $M_1\la0.8$\msun.
It is also possible to calculate $K_2$ as a function of $q$ for a
given system inclination and WD mass. For the constraints on $K_2$
and $q$ in Fig.~\ref{f:K2} and our upper estimate of $M_1=0.8$\msun,
this would suggest $i>85$\degr: implausible for a non-eclipsing
system. At a lower limit of $M_1=0.55$\msun, i.e. $q=1$, the
inclination would have to be about 65\degr for the system to lie on
the dot-dashed line in Fig.~\ref{f:K2}. While these masses and
inclinations are pure estimates, it would seem that an inclination
of $>65\degr$ and primary mass of $0.55<M_1<0.8$\msun\ are sensible
limits for V3885~Sgr.

We can briefly compare the system parameters of V3885~Sgr and
IX~Vel. BT90 derive an inclination $i=60\pm5\degr$ for IX~Vel. The
difference in the phase coverage of the narrow emission in the two
systems could point to a higher inclination for V3885 Sgr, which
would be supported by its higher $K_1$-velocity (166\kms\ vs.
138\kms).  However, it is equally plausible that the irradiated
region of the secondary is smaller in V3885~Sgr than in IX~Vel,
which would give the same effect.
%Given their spectroscopic similarity in the ultraviolet and their relative distances (IX~Vel,
%$96^{+11}_{-8}\,$pc; V3885 Sgr, $110^{+30}_{-20}\,$pc)
\citet{02hartley} argue that, given their similar UV spectra and
their relative distances, the systems are of similar inclinations,
but that IX~Vel could be the more luminous by a factor of up to two.
Assuming that the luminosity is due to accretion, then the mass
accretion rate in IX~Vel could be one to two times higher than in
V3885 Sgr. This would result in a brighter or more extended
irradiated region on the secondary. If the irradiated region were
smaller in V3885 Sgr this could explain why V3885~Sgr has a larger
ratio of $K_1/K'_2=166/96$ than IX~Vel ($K_1/K'_2=138/108$, BT90);
the smaller the irradiated region, the larger the difference between
$K_2$ and $K'_2$.

There is scant evidence in the literature to favour any particular
inclination for V3885~Sgr and little agreement between studies:
\citet{77cowley} saw no evidence for eclipsing, whereas
\citet{85haug} thought they did. Acquiring good continuum light
curves of V3885~Sgr is now a priority, both for the sake of being
better able to compare V3885~Sgr to similar systems and because a
reliable measurement of the inclination will allow us to better
understand the how disk geometry affects the irradiation of the
secondary star.

%%%%%%%%%%%%%%%%%%%%%%%%%%%%%%%%%%%%%%%%
\subsection{Spiral structure in accretion disks}
In V3885~Sgr we have seen the first conclusive evidence for
persistent spiral structure in the disk of a novalike variable.
Interestingly, the spiral waves are apparent only in H\I\ and He\I,
but not in the higher-ionisation He\II\ and N\III--C\III--C\IV\
lines. This is at odds with observations of dwarf novae in which the
spirals are typically also detected in He\IIl4686: IP Peg
\citep{99har}, WZ Sge \citep{02kuulkers}, SS Cyg \citep{96steeghs}
and U Gem \citep{01groot}. In our data the He\II\ and
N\III--C\III--C\IV lines appear to be wind-formed, so any disk
structure in these features has been submerged beneath the wind emission.

Both theory and simulation tell us that a steady state, viscous
accretion disk will not be axially symmetric about the accreting
star. For the disk to reach a steady state it must be large enough
for tidal removal of angular momentum to occur. In other words, the
disc much be large enough to be perturbed by the gravitational field
of the secondary star. We know \citep[from, e.g.][]{77pac} that the
magnetic field of the secondary will impose an elliptical
perturbation on the disk. The orientation of this perturbation is
fixed in the binary frame, making this the best frame in which to
imagine the behaviour of the gas. As gas passes around the primary
it experiences two periods of compression and two of rarefaction.
The compression can be dramatic enough as to cause the flow in the
outermost regions of the disk to shock. The vertical extent of the
outer disk will also vary quite dramatically as the gas alternately
compresses and expands (again this variation will be stationary in
the binary frame).

That elliptical variations should occur in the outer regions of CV
disks in the high state is uncontroversial. There is less agreement
as to the degree to which these perturbations propagate into the
inner disk as spiral shocks, perhaps carrying significant net
(negative) angular momentum with them. Until now, observable
elliptical disk structure has been primarily associated with dwarf
novae in outburst. In these systems however, the large variations in
disk structure (radius, thermodynamic properties, `viscosity' etc.)
that occur over the course of an outburst make it difficult to
isolate the importance of any spiral structure. In a novalike
variable, however, the disk is both large and stable, so these
should be the best systems in which to analyse spiral shocks.

The lack of identifications of elliptical disk variations in
novalike variables has been disappointing.  Although Doppler maps
for at least sixteen novalike variables have been published,
elliptical disk structure has been sighted (tentatively) only in the
eclipsing system V347~Pup \citep{98still}. It has been argued that
in novalikes any signatures from the disk would be obscured by an
outflow. However, our data show that while wind emission appears to
be obscuring spiral structure (if it is present) in the He\II\ and
N\III--C\III--C\IV\ lines, the H\I\ and He\I\ lines appear untainted
by the outflow. Therefore, the absence of disk signatures from
novalike systems, while they are detected in dwarf novae remains
puzzling. A lack of good quality spectra could be responsible, and a
re-examination of particularly the bright novalike variables now
seems justified.

V3885 Sgr is so bright that it holds the prospect of being the one
of the best candidates for studying the importance of
non-axisymmetric accretion disk structure, and higher resolution
observations are both feasible and called for. Our time and
wavelength resolution are not quite sufficient for us to either
measure the pitch angle of the spiral structure (how tightly wrapped
the spirals are), or estimate the disk radius to which it
penetrates. The former measurement would give us the Mach number of
the flow in the outer disk whilst the latter would provide a measure
of the local dissipation.

An important consideration in interpreting these results lies in
relating the emission regions to the shocks that underly them. In
particular, it is not clear how the energy released by the shocks
(presumably near the disc midplane where densities are highest) finds
its way to the upper atmosphere of the disc where emission lines are
excited. This remark applies quite generally to all observations of
disk structures.  Numerical simulations offer little insight, as they
invariably incorporate a very simple (adiabatic, polytropic or
isothermal) equation of state for the gas with little or no accounting
made of radiative transport \citep[see, e.g.][]{01boffin}.

It may not even by necessary to resort to spiral shocks. Recent
observations of spiral patterns in the disks of quiescent dwarf
novae \citep[see][and references therein]{04morales} imply that
energy dissipated by spiral shocks may not be responsible for the
observed pattern. Instead we may simply be seeing the tidally
thickened regions of the outer disk, illuminated by light from the
inner disk or boundary layer. From the point of view of V3885~Sgr
this is an attractive model, as the spirals appear in tandem with
irradiation of the secondary star. The same source of photons could
be responsible for both emission patterns, particularly as we
require the disk to be absorbing some of the boundary layer
radiation to account for the asymmetry of the secondary irradiation
pattern.

Clearly, if the spirals which we have seen in the disk of V3885 Sgr
are caused by the same mechanism as the more widely-observed spiral
structure in dwarf novae, then any explanation of this phenomenon must
be applicable to both steady-state and thermally unstable
disks. Further observations of novalike systems (the most obvious
candidate being IX~Vel) are called for in order to better understand
the three-dimensional structure of disks in binaries.
%%%%%%%%%%%%%%%%%%%%%%%%%%%%%%%%%%%%%%%%%%%%%
%%%%%%%%%%%%%%%%%%%%%%%%%%%%%%%%%%%%%%%%%%%%%

\label{lastpage}

\end{document}